# The Potential for an Innovation Winter:

## Estimating Impact of Federal Research Reductions on Faculty Activity

Robert A. Brown[1]

Boston University

February 2026

**Abstract**

The proposed reductions in federal research support proposed by the Trump Administration for 2026 would profoundly degrade the United States research universities, especially in the STEM fields and medicine (STEMM). A potentially devastating consequence would be on the funding distribution for individual faculty. Data and stochastic modeling demonstrate that the result would be large fractions of previously research active faculty having subcritical research support.

Research expenditure data from Boston University suggests that the funding distribution has a "heavy tail" — a Pareto like distribution — where a relatively small number of faculty have responsibility for a large fraction of the funding and that another fraction have minimal external support. A log normal distribution fits the expenditure data, and a multiplicative stochastic model replicates the spending distributions for the STEMM faculty and, separately, the engineering faculty. Using data for the average expenditures by 146 R1 — very research intensive — engineering schools, the model predicts that the 40 percent funding reduction proposed by the Administration would increase from 26 to 47 percent the number of R1 universities with over half their faculty with less than $100,000 of annual expenditures, if the reduction affects all R1's equally. If the most research intensive universities win a disproportionate share, this percentage of universities jumps to almost 60.

In such an environment, and in the face of other cost pressures, it would be difficult to maintain quality research and doctoral programs across many institutions, raising important questions about their optimal research strategy and organization if faced with this declining support. Ideas for navigating such a future are presented.

---

[1] Faculty of Computing and Data Sciences and College of Engineering, Boston University, Boston, MA 02215.

# Introduction

Since 1950, universities in the United States have enjoyed robust funding from the federal government for faculty-led research, resulting in a collection of research universities — essentially a system — that has been the envy of the world. Research and doctoral education in these universities, conducted alongside undergraduate and graduate professional education, feeds discoveries and innovation into the society, and is the foundation of our technical workforce.

Starting along disciplinary lines, the American research university system was built on peer review for distributing federal funding; faculty submit proposals which are reviewed by their peers, with the most meritorious receiving support. This process has democratized university research support, even though much of the funding is concentrated in a small set of world-renowned institutions; in 2022, 20 research universities received over 30 percent of the $ 55 billion of government funding. The American research university enterprise is larger than the federal support; in 2022, it totaled $ 97 billion including funding from businesses, states, and not-for-profit foundations. Actually, the second largest contributor was the universities themselves, contributing $ 24 billion, emphasizing that the system is a partnership between government and the institutions.

The increasing university support of their own research also is a sign of the precarious state of the system before the draconian decreases proposed by the Trump Administration (1,2). Dark clouds have been forming over American research universities for some time caused by an unsettlingly long list of issues; declining public confidence and support for higher education driven by increasing costs and questions of the benefits to a society with a widening income divide, declining numbers and changing demographics of college-going Americans, and the AI-powered transformation coming to higher education and every industry. To deal with these challenges research universities will need to change substantially, and it is unclear, faced with these demands, whether many can sustain, much less increase, their research support.

This was the situation as research universities approached their 75$^{th}$ birthday (set by the creation of the National Science Foundation, May 10, 1950) and before the actions of the Trump Administration created a full-fledged storm (3). If only a fraction of the Administration's proposals were to come into effect, there would be terrible damage to the research and innovation infrastructure of our country. We would be poised for an "innovation winter" that lasts decades.

As of 2025, 174 public and private universities (not counting stand-alone medical schools) are deemed R1 — very research intensive — and these institutions received 89 percent of the federal support in 2022. There are significant differences between these institutions. Data from the 2022 NSF HERD survey shows their annual research expenditures varied from 47 million to 3.8 billion dollars, including internal funding. In fact, institutional funding has been the fastest growing source for over a decade. In 2022, the R1's spent 46 cents for every dollar received from the federal government, with the ratio of internal dollars to federal support varying widely, from only 0.1 to a high of 3.2.

Funding levels are only one measure of the variation of university research environments, where administrative and laboratory support, faculty teaching loads, and internal doctoral student funding differentiate programs and institutions. One important distinction is the composition of the faculty. In some academic units the faculty is composed solely of teacher-scholars who are expected to be research-active and teach, and who are judged for promotion and tenure against standards for both efforts. In others the faculty is split between research-active, tenure-track members and those focused solely on teaching, who may neither be identified by distinctive professorial titles or be on a tenure track.[2]

In the STEMM (Science, Technology, Engineering, Mathematics, and Medicine) fields, even without intentional differentiation, some research faculty become inactive, failing to win external funding. The reality is that for each research university there is a distribution of faculty research activity, measured by either publications, citations, doctoral students trained, or external research dollars. The discussion here focuses on this distribution of research expenditures. Although it is an imperfect metric — there are fields where research money is a poor measure of productivity — in most fields external funding is critical for support of people and laboratories.

Research universities only function well when there is a critical number of research-active faculty in every academic department to sustain processes such as hiring and evaluating faculty, administering doctoral programs, and funding and supervising doctoral students. The data and modeling presented here pertains to the shape of this distribution and the impact on it of a significant decrease in external funding. Why is this analysis important now?

The partnership between American research universities and the government is crumbling, as indicated by the Administration's proposed dramatic cuts in 2026 to the research budgets of the National Science Foundation, the National Institutes of Health, and other agencies, their proposed cap on reimbursement for the indirect costs of research, the cut to the funding for graduate fellowships, decreasing support for low income students, limiting access for international students, and taxing university endowments. The list continues to grow. Taken together these actions would devastate university research leading to many institutions stepping away from research and doctoral education. For those who remain, the reductions likely will change the character of their faculty and institutions as the number of research-active faculty decreases.

---

[2] In medical schools many faculty split their time between research and clinical activity, with teaching being a minor component of their activity.

What will be the impact on the distribution of external support among the faculty? How many will remain research-active compared to the past? These questions are addressed here. The starting point is an analysis of the research activity of Boston University faculty in 2024. Two samples are considered: the entire STEMM faculty and, separately, the faculty of the College of Engineering. Examining the entire STEMM faculty demonstrates the robustness of the conclusions, while separately studying engineering connects the modeling to other universities.

The distribution of research expenditures by STEMM faculty demonstrates two important findings. First, the data are fit by a log-normal distribution with a fraction of the faculty having little research support. Second, the distribution has a "heavy tail," meaning that it exhibits for large expenditures a power law form —a Pareto distribution (4,5) — where a small number of faculty are responsible for a large percentage of the funding. These faculty either run "mega-laboratories," with extremely large research groups, or lead large, multi-faculty program grants. This distinction is important as most large projects are interdisciplinary, connecting faculty from traditional disciplines. It is essential for a modern research university to emphasize these programs.

Mathematically, the appearance of the log normal and Pareto forms are not surprising as stochastic multiplicative processes, like competition for external funding, give rise to such distributions. For example, assume that the research expenditure for the i-th faculty member changes annually (in time t) according to the simple multiplicative law:

$$S_{t+1}^i = \gamma_t S_t^i, \qquad 0 < \gamma < \infty \tag{1}$$

where $\gamma_t$ is a stochastic growth rate parameter that scales the growth or decline of expenditures over time (assumed here to be the same for all faculty). It is well known that the stochastic process, eq. (1), gives rise to both the log normal distribution (5,6) and, under certain conditions, to the power law form for S >> 1 (7,8).

A stochastic differential equation (SDE) is presented for modeling research spending by faculty within an institution. Fitting the model's parameters replicates the BU distributions and connects the model to the university's research environment. Each research university has a distribution and the sum of these make up the system. The analysis is connected to this system by data for engineering colleges in R1 universities from the USN&WR survey. Using their institutional averages, and reasonable assumptions, the model is used to estimate the impact of the Administration's proposed funding cuts on universities. Finally, strategies are discussed for mitigating the negative impact of such large reductions.

## Distributions of Research Spending

### University STEMM Faculty Data

The Boston University data is for the research expenditures in fiscal year 2024 of 911 STEMM faculty investigators across the university and includes both direct and indirect expenditures attributed to the researchers. The annual expenditures $S_i$ for the i-th investigator ranged between zero and $ 12 million; the mean expenditure was $ 518,000 and the median was $212,000. This large difference is the first indication of the distribution's skew to high values of $S_i$. Using the technique described in the Appendix, the complementary cumulative probability

distribution (ccdf) $\bar{F}(S)$ is computed and plotted in Fig. 1 as $\ln \bar{F}$ versus $\ln S$. The linear region for $S \gg 1$ has the form of the ccdf for a Pareto distribution:

$$\bar{F}(S) = k/(S)^{\alpha-1}, S \gg 1, \qquad (2)$$

with constant k, and order parameter $\alpha$. Linear regression gives $\alpha = 2.71$ with $R^2 = 0.99$. The fit is shown in Fig. 1.

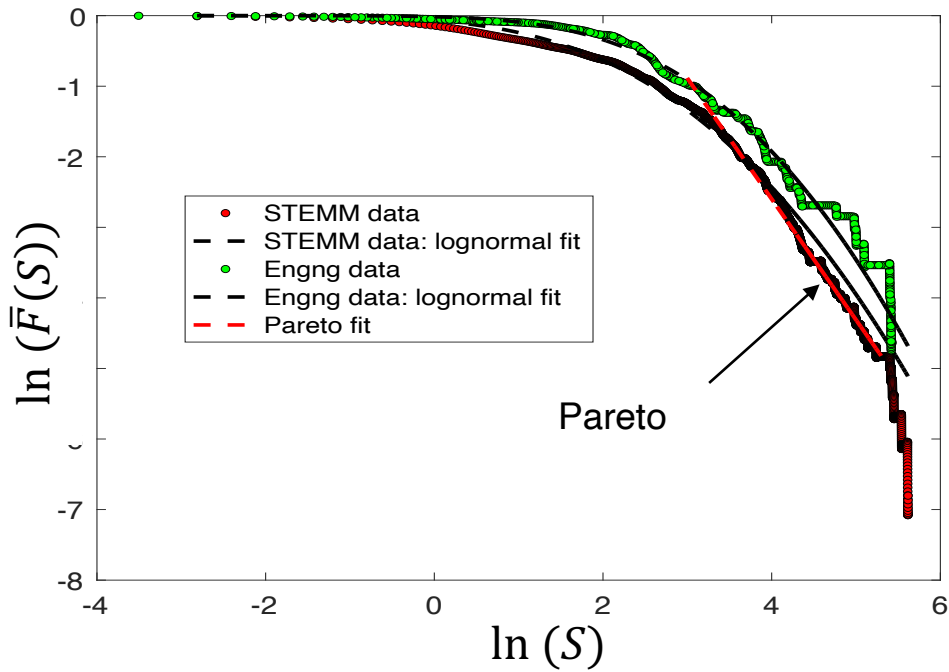

Figure 1. The complementary cumulative distribution function (ccdf) plotted versus ln(S) for both the STEMM and College of Engineering faculty from Boston University. The fits to log normal distributions and the Pareto linear regression for the STEMM faculty also are shown.

The cumulative distribution $F(S) \equiv 1 - \bar{F}(S)$, is the fraction of investigators with expenditures equal or below $S$; from the data, F(4) = 0.35, F(20) = 0.71 and $F(40) = 0.87$. An important observation is that 35 percent of the faculty had less than $100,000 (S=4) of research expenditures (both direct and indirect) in 2024. Considering present costs of funding graduate

students and post-doctoral researchers, this suggests that over a third of the faculty are not research active, measured in terms of their level of external funding.

The heavy tail of the distribution demonstrates the importance of a few faculty in the research enterprise; 87 percent of the faculty account for only 46 percent of the research funding and 13 percent account for the remaining 54 percent. In fact, two percent of the faculty are responsible for 20 percent of the funding. The large research expenditures in the tail ($S \gg 1$) represent principal investigators (PIs) with either extremely large laboratories or who lead program grants. Although these awards support other faculty who appear elsewhere in the distribution, here all expenditures are assigned to the PI.

The distribution flattens for small $S$ (small research funding), unlike a Pareto distribution; it has the form of a log normal distribution defined by the mean and variance, $(\hat{\mu}, \hat{\sigma}^2)$ for the associated Gaussian distribution. Fitting the STEMM data to the log-normal distribution as $\mathcal{N}(\hat{\mu}, \hat{\sigma}^2) = \mathcal{N}(2.1, 1.96)$ gives a reasonable approximation over the entire range of $S$. This fit to the ccdf also is shown in Fig. 1 and a comparison between the data and log-normal predictions are listed in Table 1.

**College of Engineering**

Data for 102 faculty members of the Boston University College of Engineering was analyzed separately using the same methodology; this ccdf also is included in Fig. 1. The engineering faculty have larger expenditures with mean of $766,000 and median of $346,000. Again, the distribution exhibits a tail, although the smaller sample size results in significant noise. Fitting to a log normal distribution yields the parameters $\mathcal{N}(\hat{\mu}, \hat{\sigma}^2) = \mathcal{N}(2.69, 1.54)$, shown in Fig. 1.

**Institutional Data for Colleges of Engineering**

The USN&WR survey of national universities reports the average expenditures per faculty member for engineering schools; the data used here is for the 146 Carnegie R1 universities with engineering schools for which annual expenditures per faculty ranged from $112,000 to $1,407,000, with the mean of $502,000. These expenditures are from all sources, including federal, state, business, foundations and the university, and thus are not directly comparable to the BU data which includes only external funding.

The ccdf constructed for the engineering college data is shown in Fig. 2 fit with a smoothed cubic spline approximation; the computed pdf is shown as an insert.

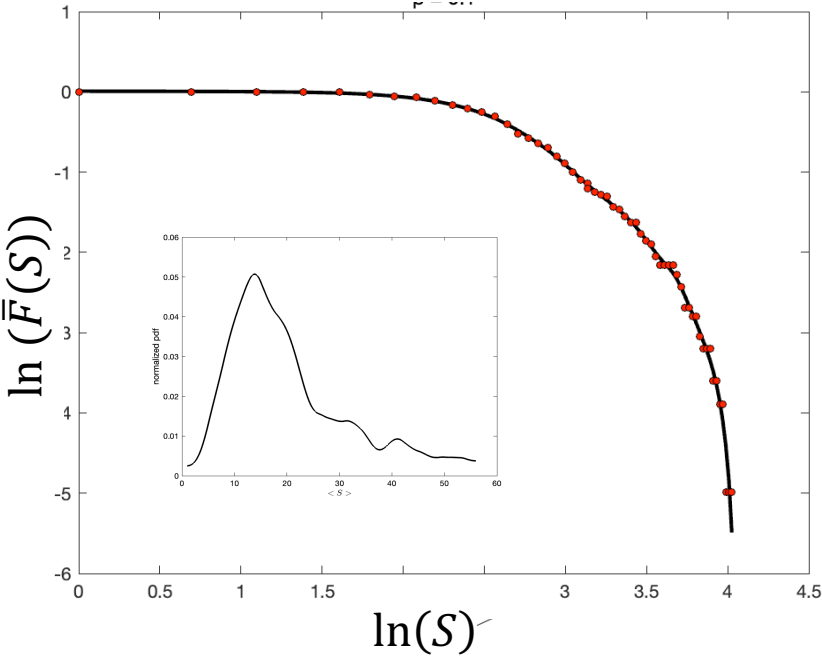

Figure 2. Complementary cumulative distribution function (ccdf) computed for the mean faculty expenditures of the R1 research universities. The pdf computed by differentiation of the smoothed ccdf is shown in the insert.

The school-level data is reasonably approximated by a log normal distribution but does not exhibit the linear tail for ($S \gg 1$). The data exhibits weak secondary peaks suggestive of a multimodal distribution.

**Results of Stochastic Differential Equation Model of Faculty Funding**

The log-normal and Pareto distributions arise empirically from the research expenditure data and also from stochastic models of a multiplicative growth process, such as eq.(1). Here a multiplicative SDE model for the annual research funding per faculty member $S^i(t)$, for the i-th institution is written as:

$$\frac{dS^i}{dt} = \gamma(S^i, t) S^i(t), \qquad (3)$$

where the stochastic growth rate parameter $\gamma(S^i, t) = \hat{\gamma}(S^i, t) + \Gamma(t)$ includes additive Gaussian white noise $\Gamma(t)$ with zero mean and covariance $\langle \Gamma(t)\Gamma(t') \rangle = 2D\delta(t-t')$; $D$ is the diffusion coefficient and $\delta(t)$ is the Dirac delta function.

The model for the growth rate $\gamma(S^i, t)$ is described in the Appendix and includes both constant and variable components proportional to $S^i$. Five parameters describe $\gamma$: $v_o^i$ is the rate of acquisition of external funding scaling both the constant and variable components; $\lambda^i$ scales the constant component; $\varepsilon^i$ scales the variable component; $\bar{S}^i$ sets the value of $S^i$ where the variable rate begins to decrease with increasing $S^i$, modeling the declining capacity of the faculty to acquire additional funding and supervise the research; and $C_o^i$ is the constant rate for internal research support.

As described in the Appendix, the stationary Fokker-Planck Kolmogorov Equation (FPKE) corresponding to eq. (3) is solved for the probability distribution function (pdf) for the distribution of faculty funding $f_{st}^i(S)$. The pdf $f_{st}^i(S)$ is illustrated in Fig. 3 using a base case defined by the parameter set $\left(\lambda^i = 0.5,\ \varepsilon^i = 1.5,\ \bar{S}^i = 10, C_o^i = 1, D = 0.05,\ v_n = 1\right)$ and $0.6 \leq v_o^i \leq 1.1$. Each distributions is similar to a log-normal distribution. The corresponding ccdf's, $\bar{F}(S)$, are shown in Fig. 3 as an insert and represent the fraction of the faculty with research expenditures exceeding $S$. For S = 10 ($250,000 of expenditure) this fraction is approximately 0.3 for $v_o^i = 0.3$ and 0.9 for $v_o^i = 1.0$. The heavy tail is visible; $\ln(\bar{F}(S))$ is approximately linear with $\ln(S)$ for $S \gg 1$. Additional features of the distribution are described in the Appendix.

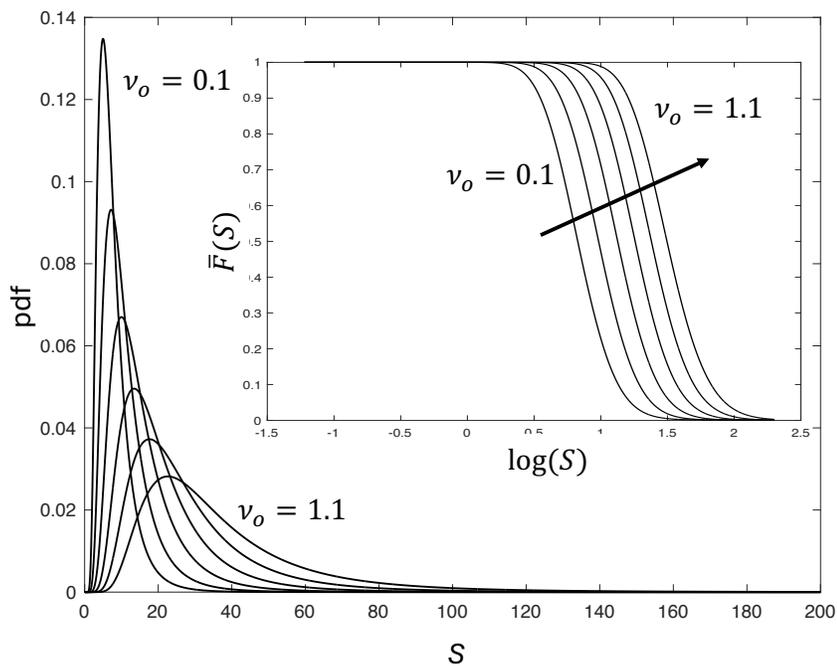

Figure 3. Examples of the pdf and ccdf predicted from the FPKE, eq. (7) for the base case parameters listed in the text any varying $v_o$ between 0.2 an 1.1 in increments of 0.1.

**Fitting BU Distributions to SDE**

The data sets for BU faculty were fit to the model pdf, eq. (A4), by varying $v_o$ and $\bar{S}$ to minimize the square residual between the model and the data. The resulting ccdf's and the data are shown in Fig. 4; the predictions for key values are listed in Table 1 and demonstrate reasonable agreement. The larger research expenditures of the engineering faculty are modelled by a larger value of $\bar{S}$ indicating a higher overall capacity for research; the values of $v_o$ are similar in both distributions. The larger deviations of the model from the data in the distribution's tail may be due to the small faculty numbers in the data.

|  | STEMM Faculty | | | Engineering Faculty | | |
| --- | --- | --- | --- | --- | --- | --- |
|  | (A) | (B) | (C) | (D) | (E) | (F) |
| Variable | Data | Log-Normal Distribution: $\mathcal{N}(2.1, 1.96)$ | SDE Model | Data | Log-Normal Distribution: $\mathcal{N}(2.69, 1.54)$ | SDE Model |
| Mean | 20.8 | 21.8 | 23.2 | 30.6 | 31.8 | 30.5 |
| Median | 8.50 | 8.17 | 10.1 | 13.9 | 14.7 | 15.5 |
| $\bar{F}(4)$ | 0.65 | 0.69 | 0.75 | 0.87 | 0.85 | 0.84 |
| $\bar{F}(10)$ | 0.47 | 0.44 | 0.50 | 0.68 | 0.62 | 0.58 |
| $\bar{F}(20)$ | 0.29 | 0.26 | 0.30 | 0.38 | 0.40 | 0.42 |
| $\bar{F}(40)$ | 0.13 | 0.13 | 0.14 | 0.23 | 0.21 | 0.22 |

Table 1. Values of merit computed for BU STEMM and engineering faculty. Shown are data (A&D), log-normal fit (B&E), and (C) Model predictions (C&F).

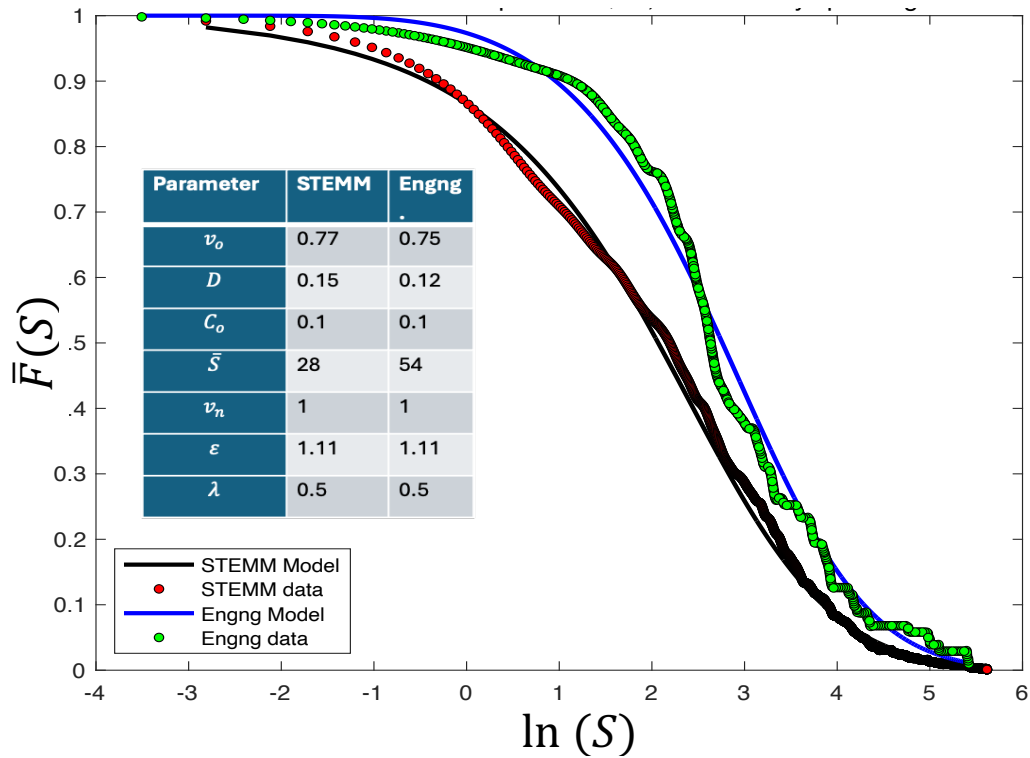

Figure 4. Regression of ccdf for STEMM and engineering faculty to SDE model; both the data and fit are shown for the parameters listed in the figure.

**Estimates for Effects of Federal Budget Reductions**

The Trump Administration proposed a 40 percent cut in university research. Although Congress has rejected his proposal, it is instructive to understand the impacts of the large reduction by analyzing the dynamics of the competition for the now smaller pool of resources. Assuming that the shape of the funding distribution for a university is set by the abilities of the research faculty and their institutional support, the increased competition for the diminished resources corresponds to decreasing $\{v_o^i\}$, the success at attracting external awards.

For engineering schools, the mean funding values multiplied by the faculty size gives the total funding per institution and the sum across institutions defines the total available resources $S_{tot}$. A reduction by the fraction $\delta$ leaves $(1-\delta)S_{tot}$. The question is how the reduction is distributed among the universities. Three methods (I-III) are examined for predicting the redistribution. In each, all institutions are assumed to have the *same distribution* as the BU engineering faculty, essentially assuming that the faculty abilities and the research environment at BU are indicative of the system. Importantly, the research environment will degrade with reductions in either indirect cost recovery or direct research support. In the model, these effects would result in lowering values of $\{\varepsilon^i, \bar{S}^i\}$, shifting the distributions to lower $S$. The calculations below bound these effects.

Methods I and II use the simple assumption that the average funding $\langle S^i \rangle$ per institution decreases in proportion to $(1-\delta)$, so that its mean decreases from the value before the cuts, $\langle S^i_{old} \rangle$, to $\langle S^i_{reduced} \rangle = (1-\delta)\langle S^i_{old} \rangle$. How this reduction affects the funding for individual faculty is approximated using the shape of the distribution. The variation in outcomes accessible by this approach is shown in Fig. 5 for $\bar{F}(S)$. Method I assumes that the redistribution is modeled by the rate of acquisition $v_o^i$ and method II uses $\bar{S}^i$, a parameter mimicking faculty capacity. Method III is a "rich stay rich" scenario in which universities with the highest $\langle S^i \rangle$ continue to be more successful at attracting external support and see a smaller overall funding decrease. It is assumed that the university with the largest value of $\langle S^i \rangle$ sees only a 20 percent decrease and a linear scaling distributes the remaining decreases to the rest of the R1's to sum the system-wide decrease to 40%. Details of the calculations are described in the Appendix.

The predictions of the three methods are summarized in Fig. 6 which shows the values of $\bar{F}(4)$ before and after a 40% reduction for each institution identified by an index compared to the funding before the reduction. Modeling the effect of the reduction by decreasing $v_o$ (Method I) predicts that the number of schools with at least 50 percent of their faculty becoming inactive ($\bar{F}(4) \leq 0.5$) increases from 26 to 69, 47 percent of the R1's after the reduction. Modeling the reduction by decreasing $\bar{S}$ (Method II) with $v_o = 0.8$ predicts the number of institutions with $\bar{F}(4) \leq 0.5$ increases from 26 to 36 institutions.

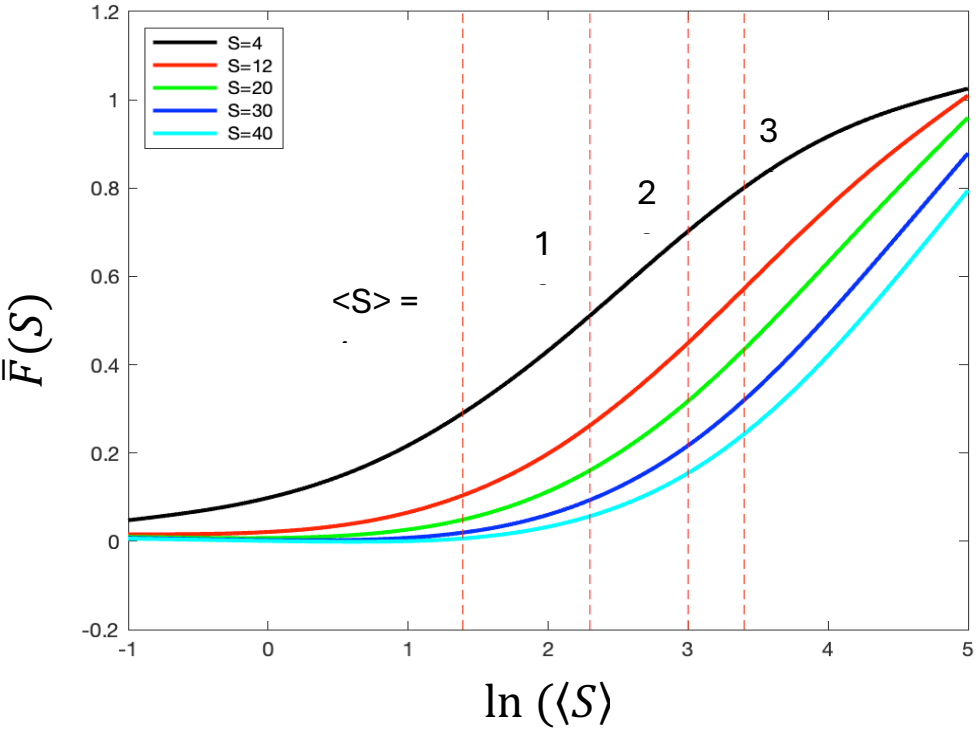

Figure 5. Representative values of $\bar{F}(S)$ computed with SDE model and BU engineering parameters; results are for varying $v_o$ in the range $0 \leq v_o \leq 1.2$. The values of $\langle S^i \rangle$ are computed from the correlation $v_o = g_{v_o}(\langle S^i \rangle)$.

When the reduction in funding is nonuniformly distributed with bias to universities with the largest expenditures per faculty (Method III), the universities with $\bar{F}(4) \leq 0.5$ increases to 82 or 56 percent of the R1's. In this scenario, the research activity in many universities would be devastated with only 50 institutions having at least a third of their faculty above the $100,000 (S=4) threshold. This result is shown on Fig. 6.

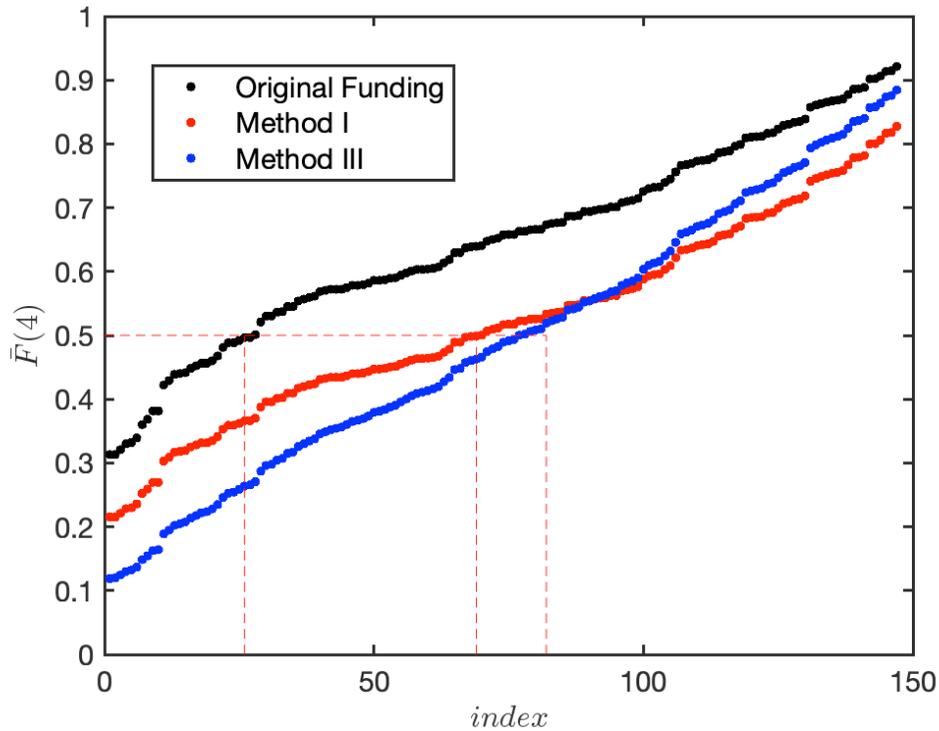

Figure 6. Predicted increase in number of faculty with less than $ 100,000 of external support caused by a 40 percent reduction in funding modelled by a uniform reduction in $v_o$.; results are shown for Methods I and III.

The potential implications of the models are clear: a reduction of the magnitude proposed by the Administration would dramatically degrade the research environment in a large number of research universities by significantly reducing the number of faculty with minimal external support . With many research universities already financially stretched, the proposed cuts, combined with other reductions, could be mortal blows, profoundly weakening these institutions and forcing some to back away from research and doctoral education.

In addition to the direct impact on faculty support, the gravest threat is the proposed cap on indirect cost (IDC) reimbursement, the funding that supports the physical and human infrastructure for research not charged directly to sponsors. The proposed 15 percent cap, if administered across all research expenditures would be greater than a factor of two reduction in IDC recovery (assuming it affects funding from all federal sponsors). An estimate for the magnitude of the cost to universities is made using data from the 2023 NSF HERD (9); the total IDC recovered by all universities was $ 17.7 billion so a fifty percent reduction is almost $ 9 billion of cost shifting to universities to be covered by tuition, endowment distributions, or state support. This is in addition to the almost $ 28 billion of research they were already supporting, including $6.8 billion on previously unrecovered IDC from externally sponsored research.

Only universities with very effective management and research strategies would emerge with some semblance of their research environment today. What might these look like? If external funding were to decrease by 40 percent, the faculty fraction with minimal research funding would grow to over half for almost half of today's R1 universities. Faculty with salaries traditionally supported by research and without significant roles in the university's educational or clinical programs — as is common in medical schools and hospitals — would be most at risk. Most probably, their numbers would decline, and this career path would become increasingly unattractive, especially in the life and biomedical sciences.

In academic units with undergraduate teaching, previously research-active faculty who lose external support would teach more and the number of tenure-track faculty members could decrease. The shift towards faculty without active research support would vary by university and field. As there are fewer research active faculty, it would become increasingly difficult for departments to maintain standards for research accomplishment by junior colleagues. It would be inevitable that more faculty would be hired solely on the teaching track and research resources would be selectively given to a chosen few who are thought to have the potential to win external funding.

In some R1's, the fraction of unfunded faculty may become so large that they cannot maintain quality faculty hiring and doctoral programs in all disciplines. Universities might move to aggregate research active investigators into special organizations where university support would be concentrated, like Germany's Max Planck Institutes buried inside existing universities. The faculty in the "heavy tail" lead programs that would be potential precursors for these units.

The most damage from the funding cuts could well be to doctoral programs that rely on external support of students. The decreased research funding coupled with the dramatic downsizing of federal fellowship programs (10) could leave many programs too small to offer quality educations. The impact on the nation would be profound as our research and development workforce shrinks. In 2023 United States institutions awarded over 36,000 doctorates in engineering, mathematics and computer science, and life and physical sciences (11); we must wonder where the funding will come from to continue this output if federal support significantly declines. The scale of this enterprise makes it unlikely that university funding could make up the loss. Of course, the students may not come either; in 2023, over 50 percent of the doctorates in engineering, computer science, and mathematics held temporary visas and it is unclear future generations of these students will be welcome or come.

**Possible Futures for American Research Universities**

This scenario paints a bleak view of the future of research universities that we can hope does not come to pass. The funding proposals by Congress for the 2026 budget may stave off the Administration's precipitous cuts but still propose decreased or stagnant funding that, compounded over time could have the same effect. We can always hope for more funding in coming years. However, "hope" is not a strategy. There are initiatives a research university can undertake now to help to winterize itself against the possibility of reductions.

Most important is to consider the research enterprise as core to the university mission and centrally organize it, including all university funding. This makes is easier to focus on quality and to make the hard decisions about the areas where the university can compete effectively for scarcer external support. The past approach of supporting "a thousand flowers blooming" likely will result in an extremely wilted research enterprise. Making these difficult choices is best done with centralized, collaborative leadership and budget management as opposed to the decentralized systems prevalent in many research universities, where dealing with revenue and expense reductions typically rests with individual academic units (3,12). Finally, it seems timely to move away from the mindset that "bigger is better" for research; everyone should understand the already large university contribution.

These suggestions are important, but they do not constitute a strategy and could result in just a smaller and less impactful version of today's research university. The larger challenge is to not simply shrink but to reinvent the institution, looking beyond the crisis to develop a vision that addresses the challenges listed at the beginning of the article, while renewing the research enterprise. An obvious dilemma would be balancing research investments against the needs for renewing undergraduate education, embracing AI, and controlling cost. Difficult choices would result; for example, it is hard to imagine universities with low graduation rates prioritizing a distressed research enterprise over student access and success.

A new vision of the research enterprise man be built around addressing society's largest challenges, such as more affordable human health care, food security, the applications of AI, and creating an inclusive and diverse democracy. This approach has the advantage of strengthening the connections of the research enterprise to the issues of the society that must support it. Having impact on these challenges requires integrating knowledge and problem solving across traditional disciplines—a greater level of academic convergence — that compels deepening collaborations in research and education.

This need points to an opportunity. It seems timely to rethink university organization to accentuate and speed convergence. Over the last two centuries academic disciplines have fostered communities for scholars and students; however, they also have promoted fiefdoms and labor cartels that make change difficult, forming what are unacceptable barriers when their institutions face a tsunami of change. It is time for universities to make disciplinary boundaries more porous for faculty and doctoral students. Faculty alignment should be balanced between disciplinary needs and research areas of convergent focus. Doctoral programs can be reenvisioned to bridge traditional disciplines, training graduates to be convergent researchers and problem solvers.

Both of these directions also align with the potential for increased corporate research support, which focuses on their mission, irrespective of academic boundaries. Corporate support will not align with all disciplines and likely cannot grow large enough to offset decreases in federal investment (according to the HERD survey in 2023, only 6 percent of university funding came from businesses compared to 55 percent from the federal government). However, increasing corporate research funding would further emphasize the relevance of research universities in the innovation ecosystem, enhancing the case for federal investment.

The same principles of convergent research and collaboration can be extended to connecting universities to create scale and impact in specific research areas, but this will require institutions to find common cause — which has been difficult in a past dominated by institutional competition for faculty, students and funding. Perhaps the environment will change.

There is no shortage of challenges facing research universities at a time when the government is working to weaken them. Even with all the gloom, there is a ray of optimism. These universities are the homes of many of the world's most talented faculty and staff. Working together with leadership they can invent a new sustainable research university, one that will be here on its 100$^{th}$ birthday.

**Acknowledgements**

The author thanks P. Lindsay for assistance with the Boston University data and K. A. Brown for his thoughtful comments on an earlier version of this manuscript.

**Declarations**

The author declares no competing interests. No funding source supported this research.

# Appendix

**Constructing ccdf for Faculty Expenditures**

To construct a ccdf from research expenditure data $\{S_i\}$, a rank plot was created in increments of $25,000 of annual research spend, expressing the probability, $P(S_i > S)$, that the expenditure of the i-th investigator exceeds S; this is the ccdf, $\bar{F}(S), S \geq 0$, computed as $\bar{F}(S_i) = \frac{1}{N}\sum_{k=1}^{N} I(S_i > k)$, where $I(X > x)$ is the index function.

**Development of SDE Model of Faculty Funding**

The growth rate parameter for faculty-led research in eq.(3) is adapted from Langevin rate equations for modeling chemical kinetics (13) and ecology (14-15). Here $\hat{\gamma}(S^i, t)$ is defined as

$$\hat{\gamma}(S^i, t) \equiv A(S^i, t) - v_n \tag{A1}$$

where $A(S^i, t)$ is the rate of incoming funding and $v_n$ is the expenditure rate; it is assumed that $v_n = 1$ thereby setting the scale for other parameters. The expression for $A(S^i, t)$ is:

$$A(S^i, t) = v_o^i \left[ \lambda^i + \frac{\varepsilon^i}{\left(1 + \frac{S^i}{\bar{S}^i}\right)} \right] + \frac{C_o^i}{S^i} \tag{A2}$$

with the five parameters $(v_o^i, \lambda^i, \varepsilon^i, \bar{S}^i, C_o^i)$. The bracketed term represents the growth rate of external funding separated into a constant rate $\lambda^i$ and a rate that scales with available funding $S^i$; the last term represents a constant rate of internal university support for research.

The variable rate component of $A(S^i, t)$ is of the form of the Beverton-Holt model for contest competition (16-17) and models a decreasing capacity for additional research funding scaled by $\bar{S}^i$. There are many rationales for including this effect, the most important being faculty capacity, as their efforts are divided between teaching, research, and service.

An alternative to integrating the SDE is to solve the associated Fokker-Planck Kolmogorov Equation (FPKE) for the probability distribution function $f^i(S, t)$:

$$\frac{\partial f^i}{\partial t} = -\frac{\partial}{\partial S}[\hat{\gamma}(S, t)S + 2DS]f^i(S, t) + \frac{\partial^2}{\partial S^2}[2DS^2 f^i(S, t)] \qquad (A3)$$

where $v_d \equiv [\hat{\gamma}(S, t) + 2DS]$ is the drift velocity and $2DS^2$ scales diffusion (18). The stationary state $f^i(S, t) \equiv f^i_{st}(S)$ is

$$f^i_{st}(S) = \frac{K_o}{2D} exp\left(\frac{1}{2D}\int_0^S \frac{\hat{\gamma}(s)}{s}ds - ln(S)\right) \qquad (A4)$$

where $K_o$ is determined so that $f^i_{st}(S)$ is normalized to unity. Equation (A4) was first derived as a general SDE describing populations in ecology (14) and is a member of the log-exponential family (19). From $f^i_{st}(S)$ other values, such as the mean or expected value, $\langle S^i \rangle$, cdf, $F^i_{st}(S)$, and ccdf, $\bar{F}^i_{st}(S) = 1 - F^i_{st}(S)$, can be computed as a function of the parameters. The integral in eq. (A4) can be expressed in closed-form, however, calculation of $K_o$ and other properties require numerical integration. The funding S is retained in dimensional form to connect to the data; a unit of S is scaled to correspond to \$25,000 of annual research expenditures; $\bar{S}$ also has units of dollars.

The heavy tail of the distribution eq. (2) is identified by taking the limit $S \gg \bar{S}$ where $v_d \to \gamma_o + 2D$, $\gamma_o \equiv v_o \lambda - v_n$. Integrating eq. (A4) gives the Pareto form $f^i_{st}(S) \sim CS^{\alpha-1}$ with $\alpha \equiv \gamma_o/2D$ and $\bar{F}(S) \sim CS^\alpha$, if $\alpha < 0$. The mathematical role of the "university contribution," $C_o$, is that in the limit $S \to 0$ where $\hat{\gamma}(S) \to C_o/S$, $f_{st}^i(S)$ approaches a constant proportional to $C_o^{-1}$, bounding $f_{st}^i(S)$ as $S \to 0$. The importance of this constant in creating the Pareto distribution was first recognized by Kersten (20); also see (21).

In eq. (A2), increasing $C_o$ models larger institutional funding that may be available in universities with larger endowments. Typically, this funding is devoted to focused research centers, giving a narrower dispersion of the funding, compared to $C_o$ which supports the entire distribution. The effect of a more focused distribution of internal funding could be included with a mixture of funding distributions of internal support, each affecting a portion of the overall distribution. This approach yields multimodal funding distributions.

The mean or expected value $\langle S^i \rangle$ is a measure of institutional research activity and increases with funding success rate $v_o^i$ and increasing $\bar{S}$ — the capacity of the faculty to lead more sponsored research. The diffusivity D also allows for significant variation. For example, decreasing $D$ tightens the pdf (Fig. A1A) around its mode and steepens the ccdf (Fig. A1B), shrinking the variation for given $v_o^i$. Alternatively, increasing $D$ widens the distribution and decreases the slope of the ccdf for large S. For given $v_o^i$, if $D \geq -\gamma_o/2$, the drift velocity is everywhere positive, spreading the distribution to larger S. The mean $\langle S^i \rangle$ increases with $D$. Also caused by the lengthening of the tail with larger $D$, the mean increases from the median, and the distribution becomes more skewed. Not surprisingly higher average research expenditures occur when there are faculty who attract large funding.

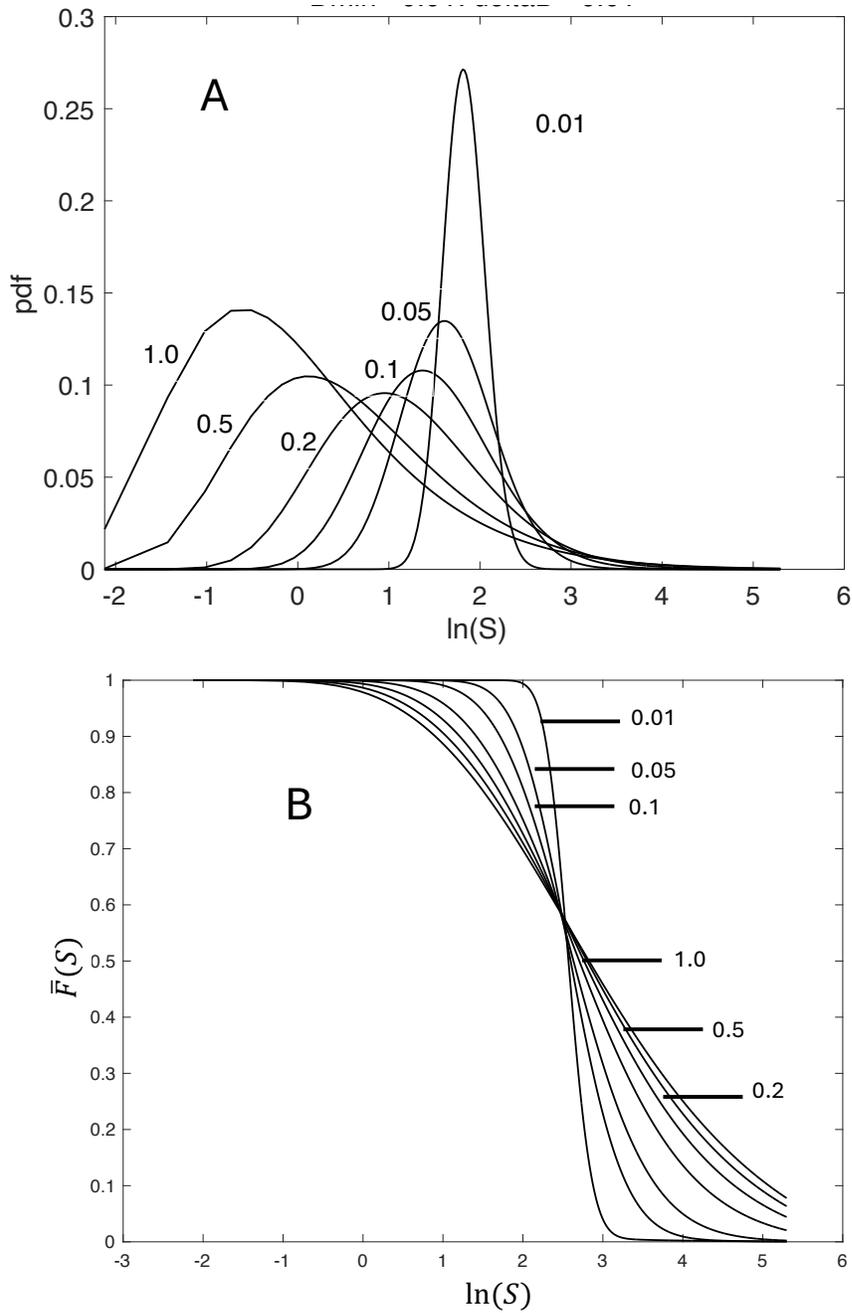

Figure A1. Effect of varying diffusivity, $0.01 \leq D \leq 1$, on the pdf (A) and ccdf (B).

**Prediction of Faculty Funding Distributions with Reduced Funding**

The following calculations were used to predict the changes in the funding distribution caused by the reduction in funding. For Method I (decreasing $v_o$), the distribution is computed using the parameters for BU engineering to produce the relationship $v_o = g_{v_o}(ln <S>)$ shown as Fig.

A2 and used to model each institution before and after the reduction. For Method II, the reduction is modelled by a decrease in $\bar{S}$ with $v_o = 0.8$ using the correlation $\bar{S} = g_{\bar{S}}(ln <S>)$ to compute the reduction. In Method III the funding reduction is nonuniformly distributed with the university with the largest value of $\langle S^i \rangle$ only a 20 percent reduction. The rest absorb the remaining reduction inversely proportional to their present research spending so that the overall decrease is 40 percent.

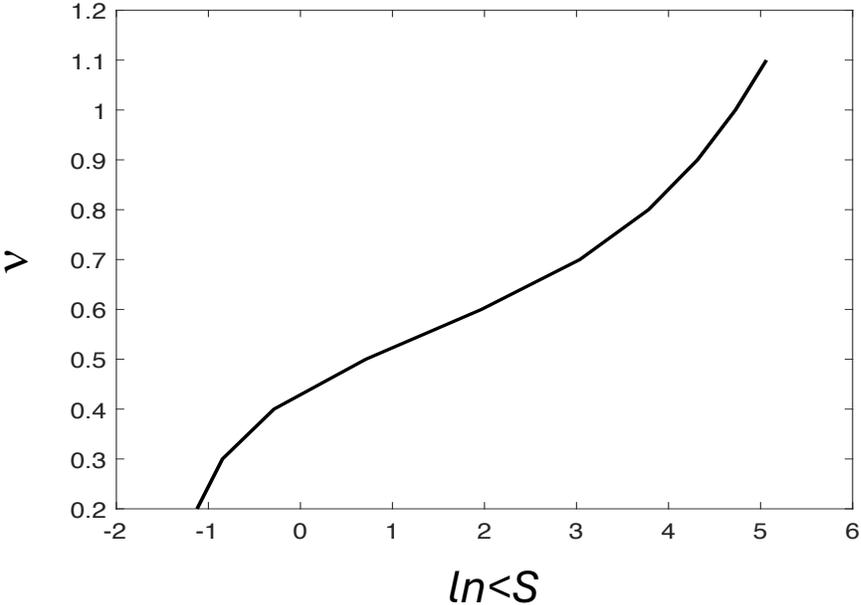

Figure A2. Correlation between $v_o$ and $ln\langle S \rangle$ used in Method II.